\documentclass[a4paper]{article}
\usepackage{hyperref}
\usepackage{INTERSPEECH2020}
\usepackage{multirow}
\usepackage{subfigure}
\usepackage{enumitem}
\newlist{steps}{enumerate}{1}
\setlist[steps, 1]{label = Step \arabic*:}
\usepackage[ruled,vlined]{algorithm2e}
\usepackage[compress]{cite}

\title{Unsupervised Learning For Sequence-to-sequence Text-to-speech For Low-resource Languages}
\name{Haitong Zhang \qquad Yue Lin}
\address{
  NetEase Games AI Lab
  }
\email{\{zhanghaitong01, gzlinyue\}@corp.netease.com}

\begin{document}

\maketitle

\begin{abstract}
Recently, sequence-to-sequence models with attention have been successfully applied in Text-to-speech (TTS). These models can generate near-human speech with a large accurately-transcribed speech corpus. However, preparing such a large data-set is both expensive and laborious. To alleviate the problem of heavy data demand, we propose a novel unsupervised pre-training mechanism in this paper. Specifically, we first use Vector-quantization Variational-Autoencoder (VQ-VAE) to extract the unsupervised linguistic units from large-scale, publicly found, and untranscribed speech. We then pre-train the sequence-to-sequence TTS model by using the $<$unsupervised linguistic units, audio$>$ pairs. Finally, we fine-tune the model with a small amount of $<$ text, audio $>$ paired data from the target speaker. As a result, both objective and subjective evaluations show that our proposed method can synthesize more intelligible and natural speech with the same amount of paired training data. Besides, we extend our proposed method to the hypothesized low-resource languages and verify the effectiveness of the method using objective evaluation.
\end{abstract}

\section{Introduction}

Sequence-to-sequence text-to-speech (S2S TTS) models consisting of an encoder-decoder-with-attention framework can generate natural speech \cite{wang2017tacotron, shen2018natural, ping2018deep, DBLP:conf/iclr/SoteloMKSKCB17, ping2018clarinet}. However, training these S2S TTS models requires tens of hour transcribed speech to produce audios with near-human naturalness. Although less data is required to produce intelligible speech, it limits overall naturalness and the model is prone to make undesirable mistakes. 

Since collecting such a large transcribed speech corpus is both expensive and laborious, researchers have started to investigate the problem of data efficiency in TTS. Some researches focused on adapting a TTS model to new speakers using a small amount of data. Some proposed to fine-tune all or parts of the pre-trained model using a small amount of data from target speakers  \cite{ chen2018sample, moss2020boffin}. Some investigated modeling speaker identities using speaker embeddings in TTS  \cite{park2019multi, nachmani2018fitting}. Some also explored a combination of speaker embeddings and fine-tuning  \cite{deng2018modeling, arik2018neural}. Some even worked on zero-shot speaker adaptation \cite{nachmani2018fitting, jia2018transfer}. 

Other researches explored building TTS model with the aid of universal data. Some studied introducing distributional textual or linguistic information into TTS within the traditional TTS paradigm  \cite{phdthesis, watts2015sentence, wang2015word}. Some  investigated training TTS models using Automatic Speech Recognition data or found data through data selection or analysis \cite{ cooper2017utterance, kuo2018data, kuo2019selection, fong2019investigating}. Recently, \cite{chung2019semi} proposed a simple yet effective semi-supervised approach to pre-train the decoder in end-to-end TTS by using only speech.  

There has been some work on data efficiency in TTS for low-resource languages. It is shown that train a multi-lingual statistical parametric speech synthesis (SPSS) model can facilitate the adaptation to new languages with a small amount of data \cite{yu2016learning, gutkin2017uniform} . A recent work \cite{chen2019end} investigated transfer learning from high-resource languages to low-resource languages. 

This work aims to alleviate the data demand for training S2S TTS by utilizing large-scale, publicly found, and untranscribed speech data. We propose an unsupervised framework for training Tacotron \cite{shen2018natural}, a state-of-the-arts S2S TTS model. Specifically, we first use Vector-quantization Variational-Autoencoder (VQ-VAE) to extract the  unsupervised linguistic units from the untranscribed  speech. We then pre-train Tacotron by using the $<$unsupervised linguistic units, audio$>$ pairs. Finally, we fine-tune the model with a small amount of $<$text, audio$>$ paired data from target speakers.


It should be noticed that our work is related to \cite{chung2019semi}. However, our work is different from \cite{chung2019semi} in several ways, constituting the main contributions of our work. The first and most significant difference is that our approach utilizes unsupervised learning to extract phone-alike linguistic units, which made it possible to pre-train the entire TTS model, while \cite{chung2019semi} separatively pre-trains each part of the model. Secondly, we also verify our approach in the hypothesized low-resource languages. Lastly, we mainly use publicly accessible data in our experiments, which can be reproduced easily.

In Section 2, we review the semi-supervised pre-training in \cite{chung2019semi} and describe our proposed unsupervised method. Section 3 details the experiment settings and results. The paper is closed with a conclusion in Section 4.

\section{Proposed Method} \label{proposed}

We use a baseline Tacotron model architecture \cite{shen2018natural}, where we use location-sensitive-attention (LSA) and phoneme sequence derived from the text. To convert the predicted spectrograms into waveforms, we use Griffin-Lim algorithm \cite{perraudin2013fast} for fast experiment cycles, since we focus on the problem of data efficiency rather than generating high-fidelity speech.
In the baseline model, the model is trained from scratch, which means all the model parameters are trained by paired data.


\subsection{Semi-supervised pre-training}

In the baseline Tacotron model, the model should simultaneously learn the textual representations, acoustic representations, and the alignment between them. \cite{chung2019semi} propose two types of model pre-training to utilize external textual and acoustic information. For textual representations, they pre-train Tacotron's encoder by the external word-vectors; for acoustic representations, they pre-train the decoder by untranscribed speech. 


\cite{chung2019semi} then fine-tune the whole model using paired data. At this step, the model focuses on learning the alignments between textual representations and acoustic ones.

\begin{enumerate}
\item[Step-1]{:}

  \begin{figure}[h]
    \centering
    \includegraphics[width =\linewidth]{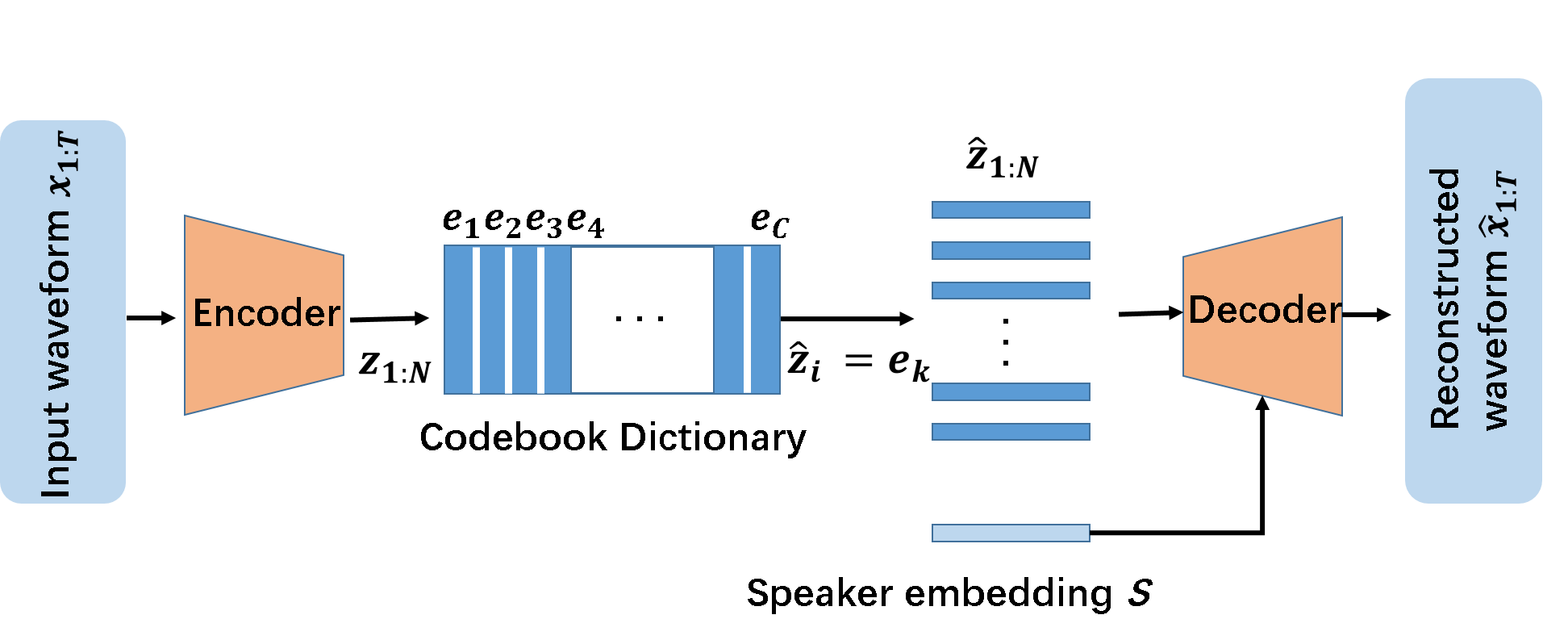}
    \caption{VQ-VAE for extracting linguistic units.}
    \label{fig:VQ}
  \end{figure}

  \item[Step-2/3]{:}

  \begin{figure}[h]
    \centering
    \includegraphics[width = \linewidth]{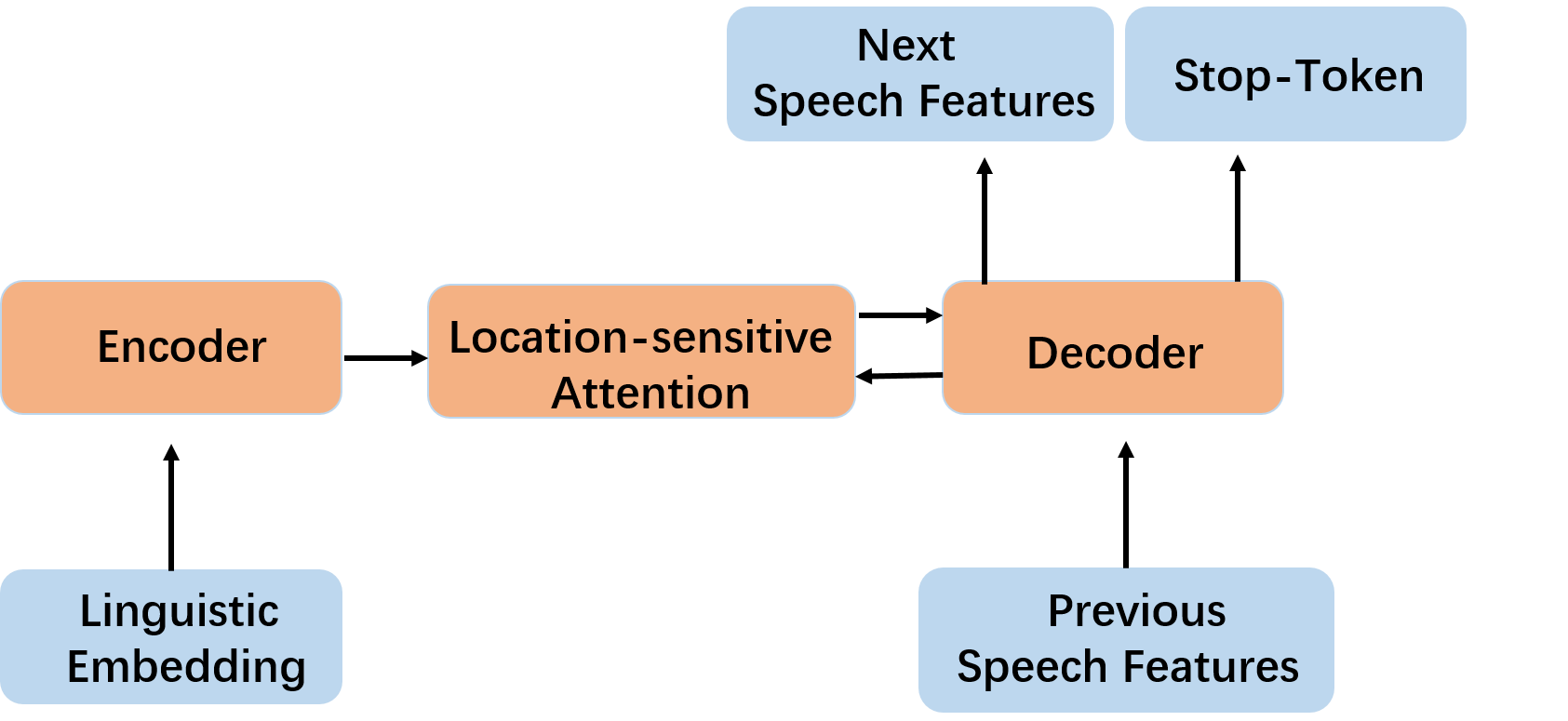}
    \caption{Tacotron model architecture studied.}
    \label{fig:Tac}
  \end{figure}

\end{enumerate}

\subsection{Unsupervised Learning for pre-training}

Although [13] shows the proposed semi-supervised pre-training helps the model synthesizes more intelligible speech, it finds that pre-training the encoder and decoder separately at the same time does not bring further improvement than only pre-training the decoder. However, there is a mismatch between pre-training only the decoder and fine-tuning the whole model. To avoid potential error introduced by this mismatch and further improve the data efficiency by using only speech, we propose to extract unsupervised linguistic units from untranscribed speech to pre-train the entire model.

Our proposed framework is provided in Algorithm \ref{algo:proposed}. The whole framework includes two models: an unsupervised model for extracting phone-alike linguistic units (see Figure \ref{fig:VQ}) and Tacotron model (see Figure \ref{fig:Tac} ).

\subsubsection{ Unsupervised linguistic units }

Unsupervised speech representation has gained a great improvement in both representation and disentangling \cite{lee2009unsupervised,  glass2012towards,hsu2017unsupervised,  van2017neural,  chung2019unsupervised, chorowski2019unsupervised}.
Among them, discretized representations are popular in language and speech community, because it is believed that language or speech is composed of a limited set of discretized units, such as characters in text and phonemes in speech. In this paper, we utilize VQ-VAE model \cite{van2017neural} as the extractor of discretized linguistic units.  

In this case, VQ-VAE acts as a recognition model similar to an automatic speech recognition (ASR) model. However, the main difference between VQ-VAE and ASR model is that VQ-VAE is trained in an unsupervised fashion while the ASR model trained in a supervised mode. This difference matters as far as low-resource languages are concerned. Whereas an ASR model for low-resource languages is not typically available, the proposed unsupervised method remains helpful in extracting linguistic units for low-resource languages.

VQ-VAE has an encoder-decoder architecture and a codebook dictionary  $e = C*D$ , where $C$ is the number of latent embeddings in the dictionary and $D$ is the dimension of each embedding. The encoder $E$ takes raw waveform $x_{1:T} =x_1, x_2, \cdots , x_T$ as inputs, and produces the encoded representation $z_{1:N} = E(x_{1:T})$ , where $N$ depends on the length $T$ and the number of down-sampling layers in the encoder. Then the continuous latent representations $z_{1:N}$ can be mapped into $\hat{z}_{1:N}$ by finding the nearest pre-defined discretized embedding in the dictionary as  $ \hat{z} = e_k$ , where $k = argmin_j$  $|| z - e_j||$, and $e_j$ is the j-th embedding in the codebook dictionary, and $j \in  {1,2, \cdots ,C}$.  Finally the latent embeddings $\hat{z}_{1:N}$ and the speaker embedding $s$ are together passed into decoder $D$ to reconstruct the raw-waveform $\hat{x} = D(\hat{z}, s)$. 

Since the model input and output are the same, the model can be trained as an auto-encoder. However, the gradients cannot be gained from the argmin operation, thus \cite{van2017neural} uses straight-through gradient estimation to approximate them. Then the final loss of the entire model is

\begin{equation}
\begin{aligned}
L     &  \;  =  \;  -log  (x \; | \; \hat{z}(x) \: , \: s) \\
      &  \;  +  \;  ||  sg(z(x)) \;  -  \; e_j) ||_2^2  \\
      &  \;  +  \;  \beta \; * \; ||  z(x)  \; - \; sg(e_j) ||_2^2 
\end{aligned}
\end{equation}

\begin{algorithm}[t]
  \caption{Proposed Method}  
  \textbf{Step1:} Training VQ-VAE using untranscribed speech
  
  \textbf{Step2:} Tacotron Pre-training:
  
  \text{2.1 Unsupervised linguistic units extraction:}
  
     \setlength{\parindent}{1em} \For{utt in untranscribed speech}{  
        \noindent1. feed utt into the trained VQ-VAE, and extract the nearest embeddings as the unsupervised linguistic units;
        
        \noindent2. delete the consecutive repeated unit from the sequence;}
        
 \noindent\text{2.2 pre-train Tacotron using $<$linguistic unit, audio$>$ pair;}
 
  \noindent\textbf{Step3:} Tacotron Fine-tuning using $<$text, audio$>$ pair.
\label{algo:proposed}
\end{algorithm}

where the first term is the negative log-likelihood to update the whole model. The second term updates the codebook dictionary, with $sg$ denotes stop-gradient operation. The third term, referred to the commitment loss, encourages the encoder output $z$ to get close to the codebook embeddings, with the hyper-parameter $\beta$ to weight the term.

\subsubsection{Tacotron Pre-training \& Fine-tuning}

After VQ-VAE is trained, we extract the unsupervised linguistic units for each utterance. We then randomly initialize an embedding table for all the unsupervised linguistic units, and the linguistic embedding sequence by looking up the table is used as the input of Tacotron. Thus, we can pre-train Tacotron by $<$linguistic embedding, audio$>$ pairs.

After the model is pre-trained as mentioned above, we fine-tune the model with some paired speech data. In the step, the inputs of the model are phoneme sequences derived from the normalized text.

\begin{table}
\caption{Result of MCD objective test of four model variants, the smaller is better. All the models are trained using 24-minute speech. The best  model (except for the upper-bound of the model) is marked in bold.}
\centering
\begin{tabular}{c c c c c}
\hline\hline

Tac       & T-Dec     & T-VQ      & T-Phone    \\ [0.5ex]
\hline

22.24 & 19.57  & \textbf{19.06}   &  18.85 &  \\ [1ex]

\hline
\end{tabular}
\label{table:MCD}
\end{table}

\begin{table}
\caption{Results of AB test of each pair of model variants. All the models are trained using 24-minute of paired data.}
\centering
\begin{tabular}{c c c c c}
\hline\hline
\multirow{2}{*}{ Model Pair} & \multicolumn{3}{c}{Preference \(\%\)} &\\


\cline{2-4}
& Former & Latter & N/A  & \\
\hline
Tac vs. T-Dec      &  1.25  & 80.25  &  18.5  &   \\
Tac vs. T-VQ       &  0     & \textbf{97.5}   &  2.5   &   \\
T-Dec vs. T-VQ     &  4.25  & \textbf{85}     &  10.75 &   \\
T-VQ vs. T-Phone   &  6.25  & 20     &  \textbf{73.5}  &   \\[1ex]

\hline
\end{tabular}
\label{table:ABtest}
\end{table}

\section{Experiment}

\subsection{Experimental setup}

We conduct experiments to show the effectiveness of our proposed method. We use the LJspeech dataset \cite{ljspeech17} for model fine-tuning. The architecture of VQ-VAE investigated in this paper is similar to \cite{chorowski2019unsupervised}. When training VQ-VAE, we use 39-dimension MFCCs as the model input. After our preliminary study, we set the codebook size into 256, and the dimension of each embedding 64. The jitter rate and $\beta$ is 0.12 and 0.25, respectively. We encourage readers to read \cite{chorowski2019unsupervised} for more details.

\cite{chung2019semi} found that 24-minute speech is the maximum amount of data that could rarely successfully build a baseline Tacotron to produce intelligible speech. Thus, in the next section, we focus on comparing all model variants trained with only 24-minute paired data.

\subsection{Results on 24-minute data}

The model variants investigated in this section include:

\begin{itemize}
    \item Tac: Tacotron trained by only LJspeech;
    \item T-Dec: Tacotron pre-trained by external speech in the semi-supervised mode, then fine-tuned by LJspeech;
    \item T-VQ: Tacotron pre-trained by external speech in the proposed mode, then fine-tuned by LJspeech;
    \item T-Phone: Tacotron pre-trained by external speech in the supervised mode, then fine-tuned by LJspeech, referred to the upper-bound of the model.
\end{itemize}

We use VCTK \cite{https://doi.org/10.7488/ds/1994} as the external speech dataset in this section. As mentioned above, we only use speech data in VCTK when pre-training T-Dec and T-VQ. For T-Phone, we use $<$text, audio$>$ paired data in VCTK for pre-training to provide the upper-bound performance in this scenario.

We conduct both objective and subjective evaluations to measure synthesis quality. For the objective evaluation, we compute Dynamic-time-warping Mel-cepstral Distortion (DTW MCD) \cite{kubichek1993mel}, which measures the distance between the synthesized and ground-truth speech, and the smaller is better. We use about 20-minute unseen speech as the evaluation data. For the subjective one, we conduct a series of AB preference tests using 20 unseen utterances of various lengths. \footnote{Speech demos are available at https://haitongzhang.github.io/DE-TTS/} 20 raters (with ten males and ten females) who are native Mandarin speakers and proficient in English are included in the subjective test.

\subsubsection{MCD objective test}

The MCD results are provided in Table \ref{table:MCD}. As in \cite{chung2019semi}, only pre-training the decoder can lower MCD. However, the proposed framework provides the best performance, whose MCD is 14.30\% lower than the baseline Tacotron. We also find that T-VQ's performance is close to that of the upper-bound of the model (i.e. T-Phone).

\begin{figure}[t]
    \centering
    \includegraphics[width=0.9\linewidth]{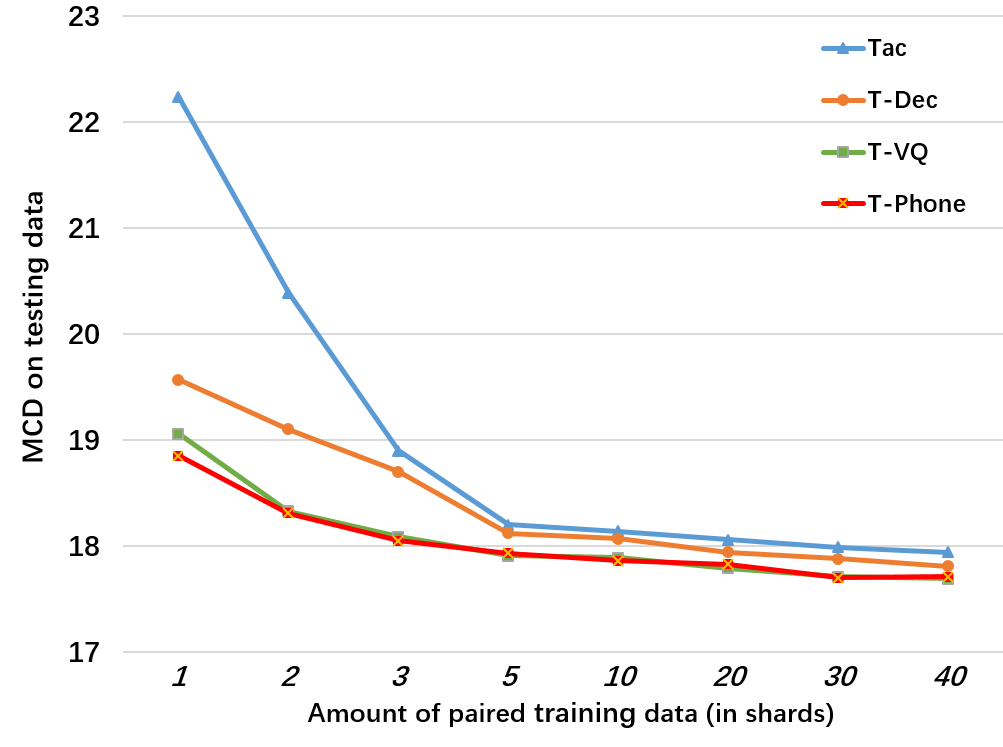}
    \caption{MCD test results of all model variants on various amount of paired data, with 1 shard equals to 24 minutes. }
    \label{fig:MCD0}
\end{figure}{}

\subsubsection{AB subjective test}

The results of the AB test are shown in Table \ref{table:ABtest}. It is clear that all pre-training techniques help to improve model performance. There is a large performance difference between baseline Tacotron and models with pre-training (i.e. T-Dec and T-VQ). We find that training the model from scratch with LJspeech can hardly get intelligible speech, partly because the quality of LJspeech is not satisfactorily high.

In the ABtest between T-Dec with the proposed T-VQ, T-VQ gets more preference from raters. From the informal listening test, we notice that synthesized speech by T-Dec is moderate in intelligibility, while T-VQ produces more intelligible speech. This indicates that pre-training by both unsupervised linguistic units and audio can further improve model performance. The reason is that in the proposed pre-training step, the model can not only learn the acoustic representation, but also the alignment between the acoustic and textual representation. Although the unsupervised linguistic embeddings are not used in fine-tuning the model, we believe that the proposed pre-training is beneficial to textual representation learning since these unsupervised linguistic units have been proven to be phone-alike \cite{chorowski2019unsupervised}. 

In the comparison between Tac-VQ and T-Phone, most raters show no preference, although raters prefer T-Phone over T-VQ by 20\%.

\subsection{Results on other amounts of data}

We also conduct MCD objective evaluation on all model variants with various amounts of data. The results are provided in Figure \ref{fig:MCD0}. Each curve in Figure \ref{fig:MCD0} represents the MCD between the ground-truth speech and synthesis by each model variant with various amounts of paired data for model training/fine-tuning. From the figure, we can see that there is a large margin between the baseline Tacotron and other model variants at 1shard (i.e. 24-minute). Another obvious trend is that with the amount of paired data increasing, the MCD differences decrease, which denotes the reducing effect of pre-training. However, regardless of the amount of paired data, T-VQ and T-Phone always achieve a lower MCD than Tac and T-Dec. 

\subsection{Results on low-resource languages}

In this section, we verify the effectiveness of the proposed approach for two hypothesized low-resource languages. In the section, we hypothesize that English and Chinese Mandarin are two low-resource languages, in which large-scale and publicly-found speech in these two languages can not be easily collected. Thus, we resort to pre-training the model by the publicly-found speech in other languages. In this section, we mainly focus on answering the two following questions:

\begin{enumerate}
    \item Is our proposed method beneficial to improving the data efficiency in this case ?
    \item What pre-training languages are more efficient in the proposed framework? Those acoustically-closely-related to target language or those acoustically dissimilar? 
\end{enumerate}

In this section, the paired data for English TTS remains LJspeech, and that for Mandarin comes from an internal news-style corpus recorded by a female speaker - Xiaomin. For training VQ-VAE and pre-training Tacotron, we use open-source corpus in the following five languages: Korean \cite{korena}, Japanese $\footnote{https://sites.google.com/site/shinnosuketakamichi/publication/jsut}$ \cite{Jap}, Spanish,  French, German$\footnote{https://voice.mozilla.org/en /datasets \label{common}}$  \cite{ardila2020common}. As mentioned above, we only use speech for training VQ-VAE and pre-training Tacotron. Only one modification is made in training VQ-VAE: the codebook size changes from $256$ to $512$, since multi-lingual data is used in this scenario. In building the English and Mandarin TTS model, we investigate the following three model variants:

\begin{itemize}
    \item Tac: Tacotron trained by paired data from LJspeech or Xiaomin;
    \item T-VQ-A: Tacotron pre-trained, in the proposed mode, by external speech in Asian languages (i.e Korean and Japanese), then fine-tuned by paired data from LJspeech or Xiaomin;
    \item T-VQ-E: Tacotron pre-trained, in the proposed mode, by external speech in European languages (i.e. Spanish, French, and German), then fine-tuned by paired data from LJspeech or Xiaomin;
\end{itemize}

To alleviate the burden of raters, we only provide MCD objective test results in this section. The MCD results of English and Mandarin TTS are provided in Table \ref{table:MCD-EN} and \ref{table:MCD-CH}, respectively. It clearly shows that our proposed pre-training approach improves the quality of the synthesized speech, which is important to low-resource languages since collecting paired data would be much more difficult. Besides, T-VQ-E out-performs than T-VQ-A in English TTS, and T-VQ-A out-performs slightly than T-VQ-E in the Mandarin experiment in most cases. This result indicates that pre-training with the speech in acoustically-close languages is more efficient than with acoustically dissimilar speech. Also, we found a similar decreasing trend in MCD with the increasing amount of fine-tuning data as in the previous section. Lastly, by comparing the best model variant in this section in English TTS (i.e. T-VQ-E) with the best model variant in the previous section (T-VQ), we found that there is still a non-negligible gap between pre-training with the speech in the target language and in the acoustically-close language (see the rows in bold in Table \ref{table:MCD-EN}), which is needed further investigation.

\begin{table}
    \caption{MCD results of model variants by various amount of paired data in English TTS. Better results are marked in bold.}
\begin{tabular}{ c c c c c c c }
\hline\hline
\multirow{2}{*}{ Model} & \multicolumn{6}{ c }{Training/Fine-tune Paired data (in shards)} \\


\cline{2-7}

 & 0.5 & 1 & 1.5  & 2 & 2.5 & 3  \\
\hline
Tac        &  28.72 & 22.24  &  21.10  & 20.39  & 19.10 & 18.90 \\
T-VQ-A     &  25.25 & 20.77  &  19.64  & 18.92  & 18.62 & 18.50 \\
T-VQ-E     &  \textbf{24.2}  & \textbf{20.14}  &  \textbf{18.73}  & \textbf{18.54}  & \textbf{18.56} & \textbf{18.45} \\
\hline
T-VQ       &  -     &  \textbf{19.06}  &  -  & \textbf{18.33}   & - &  \textbf{18.09}  \\

\hline
\end{tabular}
\label{table:MCD-EN}
\end{table}

\begin{table}
    \caption{MCD results of model variants by various amount of paired data in Mandarin TTS. Better results are marked in bold.}
\begin{tabular}{ c c c c c c c }
\hline\hline
\multirow{2}{*}{ Model} & \multicolumn{6}{ c }{Training/Fine-tune Paired data (in shards)} \\


\cline{2-7}
 & 0.5 & 1 & 1.5  & 2 & 2.5 & 3  \\
\hline
Tac        &  24.18 & 23.55  &  22.59  & 21.67  & 20.13 & 19.73  \\
T-VQ-A     &  \textbf{23.48} & \textbf{18.44}  &  16.91 & 16.31  & \textbf{15.81} & \textbf{15.49}  \\
T-VQ-E     &  23.69  & 18.63  &  \textbf{16.81}  & \textbf{16.29}  & 16.02 & 15.93  \\

\hline
\end{tabular}
\label{table:MCD-CH}
\end{table}

\section{Conclusion}
In this paper, we propose using unsupervised learning for improving data efficiency in sequence-to-sequence TTS for low-resource languages. Our method utilizes large-scale untranscribed speech to externally provide textual and acoustic information to Tacotron. We have shown that our proposed approach works in sequence-to-sequence TTS framework. Specifically, with the proposed pre-training method, Tacotron can produce intelligible speech with less paired training data. Although we conduct our experiments using Tacotron architecture, we believe that our proposed framework should be feasible in other sequence-to-sequence TTS models. We also verify the effectiveness of the method on two hypothesized low-resource languages. This promisingly indicates that even with the non-target untranscribed speech, our proposed approach could provide a significant performance improvement. Although we use hypothesized low-resource languages, we believe that our method can generalize to real low-resource languages. This significant result also sheds light on data collection for both mono-language and multi-language TTS.


Although promising results are given, there is a lot to be investigated. For example, there are many other unsupervised models to be investigated. Besides that, since we focus on the validation of the effectiveness of the proposed framework, we use Griffin-Lim as the spectrogram-to-waveform algorithm. To fully realize small paired-data sequence-to-sequence TTS, we need to investigate the adaptation of neural vocoders using small data.


\bibliographystyle{IEEEtran}
\bibliography{dataEfficiency}

\end{document}